\documentstyle[12pt]{article}

\begin{document}
\input{psfig.sty}
\begin{flushright}
\baselineskip=12pt
MADPH-99-1109 \\
\end{flushright}

\begin{center}
\vglue 1.5cm
{\Large\bf  Phenomenology Remarks in M-theory on $S^1/Z_2$\\}
\vglue 2.0cm
{\Large  Tianjun Li}
\vglue 1cm
\begin{flushleft}
Department of Physics, University of Wisconsin, Madison, WI 53706
\end{flushleft}
\end{center}

\vglue 1.5cm
\begin{abstract}
In the simplest compactification, 
we discuss the intermediate unification in 
M-theory on $S^1/Z_2$, and point out that
we can push the eleven dimension Planck scale to the
TeV range if the gauge coupling in the hidden sector is
super weak, and the particles in the hidden sector might become 
candidates of  dark matter. We also discuss the soft terms in  
non-standard embedding.
To the next leading order, we compactify the perfect square
and calculate the gravitino mass. Furthermore, we give the
general K\"ahler potential, gauge kinetic function and superpotential
if the next order correction is very large, i. e., ${{\alpha T}\over S}$
is close to 1.

\end{abstract}

\vspace{0.5cm}
\begin{flushleft}
\baselineskip=12pt
March 1999\\
\end{flushleft}
\newpage
\setcounter{page}{1}
\pagestyle{plain}
\baselineskip=14pt

\section{Introduction}

M-theory on $S^1/Z_2$ suggested by Horava and 
Witten~\cite{HW} 
is a 11-dimensional Supergravity theory with two boundaries where the
two $E_8$  Yang-Mills fields live on respectively. 
Many  studies in M-theory compactification and 
its phenomenology
implications~\cite{ Witten,Horava,BD,KC,AQ,AQR,EDCJ,VK,LLN,TIAN,BL,
JUN,JUNJUN,HLYLZH, NOY,YNOY, LT,LOD,LOW,CKM,BKL,EFN,DM,KBENAKLI,
KARIM,KARIMB,SST,NSLPT,NSLOW, LOSW, JELPP, DVN, CIMUNOZ, JEDVN,
DLOW}
 suggest that this might be the good candidate
 for super unification. In addition,  M-theory GUT Model was also
built recently~\cite{DLOW}, so, it is possible to believe that we can
make this theory more realistic in the future:
 constructing realistic GUT model in detail
  and comparing the low energy
 phenomenology with our future experiment at LHC and LEP. 

Recently, more and more people discuss the low energy gravity 
by large extra dimension and TeV string scale
~\cite{EXTRAD, CPBIQ, GKLNDT}, and their phenomenology
in the current and future colliders. The intermediate unification
has also been discussed . There are several ways to get intermediate
unification: add more particles to the model which change the RGE running
for the gauge couplings~\cite{CPBIQ},
 or some models, for example: $SU(4)\times 
SU(2)_L\times SU(2)_R$~\cite{GKLNDT},
 have the unification scale at $10^{12}$ GeV order.
We discuss the intermediate unification in M-theory on $S^1/Z_2$
~\cite{KARIMB},
using the limit $(\pi \rho)^{-1}$ is larger than $10^{-3}$ eV, we obtain 
that the low bound on the eleven dimension Planck scale and GUT scale is 
about $10^7$ GeV if the correction ${{\alpha T}\over S}$ is not
very close to -1. If we define the GUT scale as the
longest dimension in the Calabi-Yau manifold, we can push the
realistic unification scale to $10^5$ GeV, but, the eleven dimension
Planck scale is not changed. However, if ${{\alpha T}\over S}$ is 
very close to -1 in nonstandard embedding, then, we can push the
eleven dimension Planck scale  and the GUT scale to the TeV range.
The important structure for this scenario is that, 
in the hidden sector,
the gauge coupling is super weak: from about $10^{-14}$
to $10^{-30}$,   and the Calabi-Yau manifold is relatively large,
therefore, the particles in the hidden sector might play the role
of the dark matter. The major problem that
might arise in this scenario are  the
gauge unification, SUSY breaking and proton decay.

We also discuss the soft terms in the non-standard embedding
~\cite{KARIM, NSLPT}, in other
words, the gauge coupling in the hidden sector is weaker than that
in the observable sector. We discuss the soft terms in two ways:
fixing gauge couplings in the hidden sector and observable sector; 
combining the gaugino condensation with the F-term 
SUSY breaking. We find out that
comparing to the gravitino mass, 
the magnitude of $M_{1/2}$,  $A$ and $M_0$ in the non-standard embedding
is larger than 
those in the standard embedding.
 
Moreover, in the eleven dimension metric, we compactify the
perfect square to the 4-dimension, and also calculate the
gravitino mass. The gaugino condensation scale is about the order of
$10^{13}$ GeV. If we consider the gaugino condesation scale is
just after the
Calabi-Yau manifold's comfactification, the gaugino condesation scale will
be about $1.1-2.4 \times 10^{14}$ GeV, and the eleven dimension
Planck scale and the physical scale of the Calabi-Yau manifold
in the hidden sector will be about that scale, and the physical scale
of the Calabi-Yau manifold in the observable sector will be about
one half of that scale, and we can push the realistic unification
scale  to the $10^{12}$ GeV. There
is a realistic GUT model: $SU(4)\times 
SU(2)_L\times SU(2)_R$~\cite{GKLNDT}, which
 satisfies this unification scale, and it has no
problem on proton decay. We also point out that
the goldstino might be the admixture of the super partner of $S$ and $T$.

Furthermore, we give the  general k\"ahler potential,
 gauge kinetic function and superpotential in  the simplest compactification
 ~\cite{JUNJUN}.
 Because in M-theory limit, $\alpha T/S$ is  close
 to 1, we need to consider higher order terms which 
 might be large and very important
  in phenomenology analysis because it will affect the soft terms, which are
 the boundaries in running RGE.
 
\section{Intermediate Unification}
Let us review the gauge couplings, gravitational
coupling and the physical eleventh dimension
 radius in the M-theory~\cite{HLYLZH}. The relevant
11-dimensional Lagrangian is given by~\cite{HW}
\begin{eqnarray}
L_B&=&-{1\over {2 \kappa^2}}\int_{M^{11}}d^{11}x\sqrt g
R - \sum_{i=1,2}
{1\over\displaystyle 2\pi (4\pi \kappa^2)^{2\over 3}}
\int_{M^{10}_i}d^{10}x\sqrt g {1\over 4}F_{AB}^aF^{aAB} ~.~\,
\end{eqnarray}
In the 11-dimensional 
metric~\footnote{Because we think 11-dimension
metric is more fundamental than string metric
and Einstein frame, 
 our discussion in this paper use 11-dimension metric.}, 
the gauge coupling and gravitational
coupling in 4-dimension are~\cite{Witten,TIAN, HLYLZH}
~\footnote{In this paper, we do not consider the correction from 
Five-branes~\cite{NSLOW}.}:
\begin{eqnarray}
8\pi\,G_{N}^{(4)} &=& {\kappa^2 \over 
{2\pi \rho_p V_p}} ~,~ \, 
\end{eqnarray}
\begin{eqnarray}
\alpha_{\rm GUT} &=&{1\over {2 V_p (1+x)}}\,(4\pi\kappa^2 
)^{2/3} ~,~ \,
\end{eqnarray}
\begin{eqnarray}
\left[\alpha_H \right]_W &=&{1\over {2 V_p (1-x)}}\,(4\pi\kappa^2 
)^{2/3} ~,~\,
\end{eqnarray}
where $x$ is defined by:
\begin{eqnarray}
x &=& \pi^2 {\rho_p \over V_p^{2/3}} 
({\kappa \over 4 \pi })^{2/3} \int_X \omega \wedge 
{{trF \wedge F - {1\over 2} tr R \wedge R}
\over\displaystyle { 8 \pi^2}} ~,~\,
\end{eqnarray}
where $\rho_p$, $V_p$ are the physical eleventh dimension radius
and Calabi-Yau manifold volume ( which is defined by the middle 
point Calabi-Yau manifold 
volume between the observable sector and the hidden sector )
respectively, and $V_p = V e^{3\sigma}$
where $V$ is the internal Calabi-Yau volume.
From above formula, one obtains:
\begin{eqnarray}
x &=& {{\alpha_H \alpha_{GUT}^{-1} - 1} \over\displaystyle
{\alpha_H \alpha_{GUT}^{-1} + 1}} ~.~\,
\end{eqnarray}
The GUT scale $M_{GUT}$ and  the hidden sector scale
$M_H$ when 
the Calabi-Yau manifold is compactified are:
\begin{eqnarray}
M_{\rm GUT}^{-6} &=& V_p ( 1+x) ~,~\,
\end{eqnarray}
\begin{eqnarray}
M_H^{-6} &=& V_p ( 1-x) ~,~\,
\end{eqnarray}
or we can express the $M_H$ as:
\begin{eqnarray}
M_H &=& ({\alpha_H \over\displaystyle 
\alpha_{GUT}})^{1/6} M_{GUT} = ({{1+x} 
\over\displaystyle {1-x}})^{1/6} M_{GUT} ~.~\,
\end{eqnarray}
Noticing that $M_{11} = \kappa^{-2/9}$, we have
\begin{eqnarray}
M_{11} &=& \left[2 (4\pi )^{-2/3}\,  
\, \alpha_{\rm GUT}\right]^{-1/6} M_{GUT}  ~.~\,
\end{eqnarray}
And the physical 
 scale of the eleventh dimension in
the eleven-dimensional metric is:
\begin{eqnarray}
\left[\pi \rho_p\right]^{-1} &=& 
{{8 \pi}\over\displaystyle {1+x}} \left(2 
\alpha_{\rm GUT}\right)^{-3/2}
{{M_{GUT}^3}\over\displaystyle
  {M_{Pl}^2}}~,~\, 
\end{eqnarray}
where $M_{pl}=2.4\times 10^{18}$ GeV.
From the constraints that $M_{GUT}$ and $M_H$ is smaller than
the scale of $M_{11}$, one obtain:
\begin{eqnarray}    
\alpha_{GUT} \leq {{(4 \pi)^{2/3}}\over\displaystyle
2} ~;~
\alpha_H \leq {{(4 \pi)^{2/3}}\over\displaystyle
2} ~,~ \,
\end{eqnarray}
or 
\begin{eqnarray}    
\alpha_{GUT} \leq 2.7 ~;~
\alpha_H \leq 2.7 ~,~ \,
\end{eqnarray}
 For the 
standard embedding,  the upper bound on
x is 0.97 ( $x < 0.97 $ ),
for $\alpha_{GUT} = {1\over 25}$. 
From the constraints 
that $ \left[\pi \rho_p\right]^{-1} $ is smaller
than the scale of $M_{11}$, we obtain that:
\begin{eqnarray} 
M_{GUT} \alpha_{GUT}^{-2/3} \leq 
\sqrt {1+x} 2^{1/6} (4 \pi )^{-4/9}
M_{Pl} ~,~ \,
\end{eqnarray}
which is obviously satisfied in the standard embedding.
However, if we
 can consider non-standard embedding $ x < 0 $
 ~\cite{KARIM,NSLPT}, i. e., 
the gauge coupling in the observable sector is larger
than the gauge coupling in the hidden sector, the low bound on x
is:
\begin{eqnarray} 
x_{lb} \geq 2^{-1/3} (4 \pi )^{8/9} 
(\alpha_{GUT})^{-4/3} 
{{M_{GUT}^2} \over\displaystyle {M_{Pl}^2}} -1 ~.~ \,
\end{eqnarray}

Now, let us discuss the intermediate unification~\cite{KARIMB}. 
We can write the eleven dimension Planck scale and
the physical 
 scale of the eleventh dimension in terms of the $\alpha_{GUT}$
 and $M_{GUT}$:
\begin{eqnarray}
M_{11} &=&  1.18 \left[\alpha_{\rm GUT}\right]^{-1/6} M_{GUT}  ~,~\,
\end{eqnarray} 
\begin{eqnarray}
\left[\pi \rho_p\right]^{-1} &=& 
{{1.54} \over\displaystyle {1+x}} \left( 
\alpha_{\rm GUT}\right)^{-3/2}
\left({{M_{GUT}}\over\displaystyle
  {10^{16} {\rm GeV}}}\right)^{3} \times 10^{12} {\rm GeV}
~.~\, 
\end{eqnarray}
Because  the gauge coupling $\alpha_{GUT}$ in  the observable sector 
can not be too large ( at least from current model building as far as we
know ), and if the gauge coupling $\alpha_{H}$ in  the hidden sector 
was not too small, we might think that $M_{GUT}$ is the
 important factor which determines the $M_{11}$ and  
$\left[\pi \rho_p\right]^{-1}$.  

First, we consider the standard embedding, i. e. $x > 0$.
Assuming that $\alpha_{GUT}= {1\over 25}$, we can write
the eleven dimension Planck scale and
the physical 
 scale of the eleventh dimension: 
\begin{eqnarray}
M_{11} &=&  2.02 M_{GUT}  ~,~\,
\end{eqnarray} 
\begin{eqnarray}
\left[\pi \rho_p\right]^{-1} &=& 
{{1.93} \over\displaystyle {1+x}} 
\left({{M_{GUT}}\over\displaystyle
  {10^{16} {\rm GeV}}}\right)^{3} \times 10^{14} {\rm GeV}
~.~\, 
\end{eqnarray}
Noticing that in this case, $x > 0$, the key factor which will affect the
eleven dimension Planck scale and
the physical 
 scale of the eleventh dimension is $M_{GUT}$. And then we can obtain 
 different unificaton scale. Using the limit that $\left[\pi \rho_p\right]^{-1}$
 is larger than $10^{-3}$ eV, we obtain that the low bound on 
 $M_{11}$ is about $3.5\times 10^7$ GeV, and the low bound on 
 $M_{GUT}$ is about  $1.73\times 10^7$ GeV.
 
Second, we consider the non-standard embedding~\cite{KARIM, NSLPT}, i. e.,
 the gauge coupling in the 
observable sector is stronger than that in the hidden sector.
Let us assume that $\alpha_{GUT} = 0.15$,
we can write
the eleven dimension Planck scale and
the physical 
 scale of the eleventh dimension: 
\begin{eqnarray}
M_{11} &=&  1.62 M_{GUT}  ~.~\,
\end{eqnarray} 
\begin{eqnarray}
\left[\pi \rho_p\right]^{-1} &=& 
{{2.66} \over\displaystyle {1+x}} 
\left({{M_{GUT}}\over\displaystyle
  {10^{16} {\rm GeV}}}\right)^{3} \times 10^{13} {\rm GeV}
~.~\, 
\end{eqnarray}       
If we do not consider the factor ${1 \over {1+x}}$, using the limit 
that $\left[\pi \rho_p\right]^{-1}$
 is larger than  $10^{-3}$ eV, we obtain the low bound on 
 $M_{11}$ is $5.4\times 10^7$ GeV, and the low bound on 
 $M_{GUT}$ is $3.35\times 10^7$ GeV. However, we might get TeV
 unification scale or any scale unificaton ( 11-dimension
 Planck scale is at the same order) if 
 theory is consistant, i. e., it can have that scale as the
 unification scale,
 and avoid the problem like proton decay, SUSY breaking etc.,
  because we can set the $x$ very close
 to -1, but $x > x_{lb}$. For example,
 asuming that
 $\left[\pi \rho_p\right]^{-1}$
 is  $10^{-3}$ eV, $M_{11}=5.4$ TeV or $M_{GUT}=3.35$ TeV, we
 obtain that $M_H=33.5$ GeV and $\alpha_H=7.5\times 10^{-14}$,
 therefore, the physical volume of Calabi-Yau manifold 
 in the hidden sector is relative large and the gauge coupling in the
 hidden sector is super weak, but it is  much stronger than gravity. 
 And one
 might think that the particles in the hidden sector will play the role of
 dark matter.  
 In addition, using $\alpha_{GUT} = {1\over {25}}$, the above 
  argument will not change, 
asuming that
 $\left[\pi \rho_p\right]^{-1}$
 is  $10^{-3}$ eV, $M_{11}=3.5$ TeV or $M_{GUT}=1.73$ TeV, we
 obtain that $M_H=17.3$ GeV and $\alpha_H=2\times 10^{-14}$.
 In addition,  for $\alpha_{GUT}=0.15$,
 $x=x_{lb}=1.85\times 10^{-28} -1$,  with
  $M_{11}=5.4$ TeV or $M_{GUT}=3.35$ TeV, we
 obtain that $M_H=71$ MeV and $\alpha_H=1.4\times 10^{-29}$. For
$\alpha_{GUT}=0.04$,
 $x=x_{lb}=2.86\times 10^{-28} -1$,  with
  $M_{11}=3.5$ TeV or $M_{GUT}=1.73$ TeV, we
 obtain that $M_H=40$ MeV and $\alpha_H=5.7\times 10^{-30}$.    
 In short,
 $  x_{lb} < x < 10^{-12} -1$,
 and for $\alpha_{GUT}=0.04$, and  $M_{11}=3.5$ TeV or $M_{GUT}=1.73$ TeV, we have
  $33.5 GeV > M_H > 71 MeV$, $7.5\times 10^{-14} > \alpha_H > 1.4\times 10^{-29} $,
  for $\alpha_{GUT}=0.15$, and  $M_{11}=5.4$ TeV or $M_{GUT}=3.35$ TeV,
  we obtain 
  $17.3 GeV > M_H > 40 MeV$, $2\times 10^{-14} > \alpha_H > 5.7\times 10^{-30} $. 
 By the way, setting $x=x_{lb}$, we can obviously
 put the $M_{GUT}$ and $M_{11}$ 
 ( $(\pi \rho)^{-1} = M_{11}$ ) at any scale if the theory is consistant.

In the above discussion, one does not consider the detail of the
length of each dimension in the Calabi-Yau manifold. If we just consider
it as a 6-dimension manifold,  we can 
define $M_{GUT} = l^{6-n} L^n$, i. e.,
we assume that $n$ dimension have  relatively larger
 length and $6-n$ dimension have
 relatively smaller length. Assuming $l^{-1}=M_{11}$, and considering
$L^{-1}$ as realistic GUT scale $M_{GUT}^r$, we obtain that:
\begin{eqnarray}
M_{GUT}^r= {{\alpha_{GUT}^{1/n}} \over\displaystyle 1.18^{6/n}}
M_{11}
~.~\, 
\end{eqnarray}
This is another way that we can push the realistic
unification scale smaller ( the eleven dimension
scale and $M_{GUT}$ will not change), 
for n=1, $M_{GUT}^r = 0.015 M_{11}$ or $M_{GUT}^r={1\over {33.5}} M_{GUT}$
for $\alpha_{GUT} = 0.04$, 
 $M_{GUT}^r = 0.056 M_{11}$ or $M_{GUT}^r={1\over {11.13}} M_{GUT}$
for $\alpha_{GUT} = 0.15$.
And  for n=2, $M_{GUT}^r = 0.12 M_{11}$ or $M_{GUT}^r={1\over {4.1}} M_{GUT}$
for $\alpha_{GUT} = 0.04$, 
 $M_{GUT}^r = 0.24 M_{11}$ or $M_{GUT}^r={1\over {2.6}} M_{GUT}$
for $\alpha_{GUT} = 0.15$. However, using above low bound,
we just put the low bound of $M_{GUT}^r$ to the order of 
$10^5$ GeV, which is relatively
higher than the reach of LHC and LEP.

Even though we consider standard embedding where $ 0.97 > x >0$,
if we assume that we  only
observe the materials or physics at our boundary or  the observable sector,
and the fifth dimension is very large~\footnote{
Of course, it is finite, considering $M_{GUT}$= 1.73 TeV and 
$\alpha_{GUT}=0.04$ ( or $M_{GUT}$= 3.35 TeV and 
$\alpha_{GUT}=0.15$), we obtain that
$(\pi \rho)^{-1}$= $10^{-24}$ GeV or 
$\pi \rho=1.97\times 10^8$ meter.}, then $M_{GUT}$ can be
at any scale if the theory is consistant.

\section{Soft Terms in Nonstandard Embedding}

The k\"ahler potential, gauge kinetic function and
the superpotential in the simplest compactification
of M-theory on $S^1/Z_2$ are ~\cite{NOY,LOD}:
\begin{eqnarray}
K &=& \hat K + \tilde K |C|^2 ~,~ \,
\end{eqnarray}
\begin{eqnarray}
\hat K &=&  -\ln\,[S+\bar S]-3\ln\,[T+\bar T] ~,~\,
\end{eqnarray}
\begin{eqnarray}
\tilde K &=& ({3\over\displaystyle {T+\bar T}} +
{\alpha\over\displaystyle {S+\bar S}}) |C|^2  ~,~ \,
\end{eqnarray}
\begin{eqnarray}
Ref^O_{\alpha \beta} &=& Re(S + \alpha T)\, \delta_{\alpha \beta} ~,~\,
\end{eqnarray}
\begin{eqnarray}
Ref^H_{\alpha \beta} &=& Re(S - \alpha T)\, \delta_{\alpha \beta} ~,~\,
\end{eqnarray}
\begin{eqnarray}
W= d_{x y z} C^x C^y C^z ~,~\,
\end{eqnarray}
where $S$, $T$ and $C$ are dilaton, moduli and matter fields 
respectively. $\alpha$ is a next order correction constant 
which is related to the  Calabi-Yau manifold.
And the nonperturbative superpotential in the simplest model
 is~\cite{LOW}:
\begin{eqnarray}
W_{np} &=& h~ exp(-{{8 \pi^2}\over\displaystyle C_2(Q)}
(S- \alpha T)) ~,~\,
\end{eqnarray}
where the group in the hidden sector is Q which is
a subgoup of $E_8$, and
$C_2(Q)$ is the
quadratic Casimir of Q. By the way, 
$C_2(Q) = 30, 12$ for $Q$ is $E_8$ and $E_6$, respectively. 

With those information, one can easily obtain the following
soft terms\cite{JUN,CKM}:
\begin{eqnarray}
M_{1/2}&=&{{\sqrt 3  M_{3/2}} \over\displaystyle {1+x}}
(sin\theta +{x\over \sqrt 3} cos\theta ) ~,~\,
\end{eqnarray}
\begin{eqnarray}
M_0^2&=& M_{3/2}^2 - 
{{3  M_{3/2}^2} \over\displaystyle 
{(3+x)^2}}(x(6+x) sin^2\theta +
\nonumber\\&&
 ( 3 + 2x ) cos^2\theta
- 2 \sqrt 3 x ~sin\theta ~cos\theta) ~,~ \,
\end{eqnarray}
\begin{eqnarray}
A&=&- {{\sqrt 3  M_{3/2}} \over\displaystyle 
{(3+x)}} ((3-2x) sin\theta + \sqrt 3~ x ~cos\theta) ~,~ \,
\end{eqnarray}
where $M_{3/2}$ is the gravitino
mass, the quantity x defined above is the same as that in the 
last section and can also be expressed as
\begin{eqnarray}
x={{\alpha ( T + \bar T )} \over\displaystyle { S + \bar S }} 
 ~.~\,
\end{eqnarray}

If one combined the gaugino condesation scenario with above soft terms,
one obtain the angle $\theta$~\cite{JUN}:
\begin{eqnarray}
tan\theta &=& {1\over\displaystyle \sqrt 3} 
{{1+{2\pi \over\displaystyle {C_2(Q)}}(\alpha_{GUT}^{-1}
+\alpha_H^{-1})}\over\displaystyle 
{1-{2\pi \over\displaystyle {3 C_2(Q)}}(\alpha_{GUT}^{-1}
- \alpha_H^{-1})}} ~.~ \,
\end{eqnarray}
so, the soft terms $M_{1/2}, M_0,$ and $A$  are  the
functions of the gravitino mass and the
gauge coupling ( $\alpha_H$ ) in the hidden sector when one consider
the hidden sector gaugino condensation.

Now, we numerically evaluate those soft terms in nonstandard embedding
~\cite{KARIM, NSLPT}.
We take the gauge coupling in the observable sector as: $\alpha_{GUT} = 0.15$.
First, we just consider $F$ term of $S$ and $T$ SUSY  breaking but we do
not consider the gaugino condensation, we draw  the soft terms
in the unit of gravitino mass versus
$\theta$ in fig. 1, 2, 3, 4~\footnote{Constraint that
the scalar mass square $M_0^2$ should be larger than 
zero is considered.} 
 for 
$\alpha_H =0.025, 0.05, 0.075, 0.1$ respectively, we can see that when  
$\alpha_H$ is small, the magnitude of $M_{1/2}$ and $A$ are  larger
than that of  $M_0$,
and comparing to the gravitino mass, 
the magnitude of $M_{1/2}$,  $A$ and $M_0$ are relatively larger than those 
 in the standard embedding. 
Combining the gaugino condensation and the F-term SUSY breaking,
 we draw the soft term versus $\alpha_H$ in 
  fig. 4, 5 where the hidden sector group are 
 $E_8$ and $E_6$ respectively~\footnote{
 We use $E_8$ as an example, although it is not a proper subgroup of
 $E_8$.}, we obtain that, all the soft terms decrease
 when we increase the hidden sector gauge coupling, but the variation of
 $M_0$ is very small, and the variation of $M_{1/2}$ is very large at small
 $\alpha_H$.
 When $\alpha_H$ large enough, i. e., $\alpha_H > 0.1$,
  the variations of all the soft terms 
 are very small.
  Noticing that in order to obtain
  the gravitino mass at the 
 hundred GeV range,   
   the hidden sector gaugino
 condensation scale can not be too small,
i.e.,  we can not let $\alpha_H$ be too small. We 
 pick two points as representative points: $\alpha_H = 0.05$ and $0.1 $
 with $E_8$ as hidden sector gauge group. The soft terms for $\alpha_H=0.05$ 
 in the unit of gravitino mass are:
\begin{eqnarray} 
 M_0=1.23 ~;~ M_{1/2}= 2.64 ~;~ A = 2.20 ~,~ \,
\end{eqnarray} 
the soft terms for $\alpha_H=0.1$ in the units of the gravitino mass 
are: 
\begin{eqnarray} 
 M_0=1.03 ~;~ M_{1/2}= 1.85 ~;~ A = 1.80 ~.~ \,
\end{eqnarray}

\section{Compactification of Perfect Square and Gravitino Mass}
First, we would like to review Horava's result~\cite{Horava}. 
The 10-dimensional and 11-dimensional conventions are as in~\cite{HW}.
The space time signature is - + $\dots$ +. Eleven dimensional vector indices
are written as I, J, K, L, $\dots$ . The 11-dimensional $\Gamma$ matrices
are 32$\times$32 real matrices satisfying 
$\{\Gamma_I, \Gamma_J \}= 2 g_{IJ}$, with $g_{IJ}=\eta_{mn} e^m_I
e^n_J$, the eleven dimensional metric. Each boundary of 11-dimensional manifold 
surport one $E_8$ Yang-Mills supermultiplet. 
In the simplest compactification,
one of the $E_8$'s will be broken by 
the spin connection embedding to a grand-unified $E_6$ group, while the 
other $E_8$ in the hidden sector 
will not be broken.
  The adjoint index of this hidden $E_8$ will be denoted by 
$a,b,\ldots$.  

On $R^4\times S^1/ Z_2\times X$, we will use four-dimensional vector indices 
$\mu,\nu,\ldots$ that parametrize the flat Minkowski space $R^4$, and vector 
indices $i,j,k,\ldots$ and their compex conjugates $\bar i,\bar j,
\bar k,\ldots$ that correspond to a complex coordinate system on the 
Calabi-Yau three-fold $X$.  The ten-dimensional vector indices that 
parametrize $R^4\times X$ will be written as $A,B,C,\ldots$.  The other 
conventions on $X\times S^1/ Z_2$ are as in~\cite{HW}.  

The effective Lagrangian for this theory was constructed
 in~\cite{HW}.  It contains the eleven-dimensional supergravity 
multiplet $e_I{}^m,\psi_J$ and $C_{IJK}$ in the bulk, coupled to each of
two $E_8$ 
Yang-Mills supermultiplet $A_B^i,\chi^i$ at each of the two ten-dimensional 
boundaries.  

To order $\kappa^{2/3}$, the Lagrangian is given by
\begin{eqnarray} 
\cal L&=&{1 \over \kappa^2} \int_{M^{11}}d^{11}x\sqrt g
\left(-{1\over 2} R-{1\over 2} \bar\psi_I\Gamma^{IJK}D_J {{\Omega+
\hat\Omega} \over 2})\psi_K-
{1 \over 48}
G_{IJKL}G^{IJKL}
\right.\nonumber\\&&\left.
-{\sqrt 2 \over {384}}\left(\bar\psi_I\Gamma^{IJKLMN}\psi_N
+12\bar\psi^J\Gamma^{KL}\psi^M\right)\left(G_{JKLM}+\hat G_{JKLM}\right)
\right.\nonumber\\&&\left.
-{\sqrt 2 \over {3456}}\epsilon^{I_1I_2\ldots I_{11}}
C_{I_1I_2I_3}G_{I_4\ldots I_7}G_{I_8\ldots I_{11}}\right)
\nonumber\\&&
+{1 \over\displaystyle {2\pi(4\pi\kappa^2)^{2/3}}}
\int_{M^{10}_i}d^{10}x\sqrt g\left(
-{1\over 4}F^i_{AB}F^{i AB}-{1\over 2}\bar\chi^a\Gamma^AD_A(\hat\Omega)
\chi^a
\right.\nonumber\\&&\left. 
-{1 \over 8}\bar\psi_A\Gamma^{BC}\Gamma^A\left(F^i_{BC}+
\hat F^i_{BC}\right)\chi^a+{\sqrt 2 \over {48}}\left(\bar\chi^a\Gamma^{ABC}
\chi^a\right)\hat G_{ABC\,11}\right)
~,~ \,
\end{eqnarray} 
where the definitions of $\hat\Omega$, $\hat 
F^a_{AB}$ and $\hat G_{IJKL}$ can be found in~\cite{HW} and i in 
$F$, $\hat F$ and $M_i^{10}$ is boundary index, i. e., i=
 1, 2.

The relevant supersymmetry transformations of the gravitino fields are:
\begin{eqnarray}
\delta\psi_A&=&D_A\eta+{\sqrt 2 \over {288}}G_{IJKL}
\left(\Gamma_A^{IJKL}-8\delta_A^I\Gamma^{JKL}\right)\eta
\nonumber\\&&
-{1 \over {576\pi}}\left({\kappa \over {4\pi} }\right)^{2/3}
\delta(x^{11})
\left(\bar\chi^a\Gamma_{BCD}\chi^a\right)\left(\Gamma_A^{BCD}-6\delta_A^B
\Gamma^{CD}\right)\eta+\ldots
~,~ \,
\end{eqnarray}
\begin{eqnarray}
\delta\psi_{11}&=&D_{11}\eta+{\sqrt 2 \over {288}} G_{IJKL}\left(
\Gamma_{11}^{IJKL}-8\delta_{11}^I\Gamma^{JKL}\right)\eta
\nonumber\\&&
+{1 \over {576\pi}}\left({\kappa \over {4\pi}}\right)^{2/3}
\delta(x^{11})\left(\bar\chi^a\Gamma_{ABC}\chi^a\right)\Gamma^{ABC}\eta+
\ldots
~,~ \,
\end{eqnarray}
where the $\ldots$ denote terms of order $\kappa^{4/3}$, as well as 
known terms 
of order $\kappa^{2/3}$ bilinear in the gravitinos that we do not need
to use it.  

With another term at relative order of $\kappa^{4/3}$, Horava obtained
the perfect square in the M-theory on $S^1/Z_2$~\cite{Horava}:
\begin{eqnarray}
-{1\over {12\kappa^2}}\int_{M^{11}}d^{11}x\sqrt g\left(
G_{ABC 11}-{\sqrt 2\over {16\pi}}\left({\kappa \over {4\pi}}\right)^{2/3}
\delta(x^{11})\,\bar\chi^a\Gamma_{ABC}\chi^a\right)^2  
~,~ \,
\end{eqnarray}   
this perfect square is similar to that in the weakly coupled 
heterotic string,
in which one change $G_{ABC 11}$ to $H_{ABC}$.

In the eleven dimension metric and simplest
compactification, one can compactify this perfect square to
5-dimension, it is:
\begin{eqnarray}
L^{(5)}&=&-{1\over {12\kappa^2}}\int_{M^{5}}d^{5}x V_p (1-x) \sqrt g\left(
G_{ABC 11}-
\right.\nonumber\\&&\left.
{\sqrt 2\over {16\pi}}\left({\kappa \over {4\pi}}\right)^{2/3}
\delta(x^{5})\,\bar\chi^a\Gamma_{ABC}\chi^a\right)^2  
~,~ \,
\end{eqnarray} 
and to the 4-dimension, it is:
\begin{eqnarray}
L^{(4)}&=&-{1\over {12\kappa^2}}\int_{M^{4}}d^{4}x 2 \pi \rho_p V_p (1-x) 
(1-{x\over 3})
 \sqrt g\left(
G_{ABC 11}-
\right.\nonumber\\&&\left.
{1\over {1-{x\over 3}}} {1 \over {2 \pi \rho_p}}
{\sqrt 2\over {16\pi}}\left({\kappa \over {4\pi}}\right)^{2/3}
\delta(x^{5})\,\bar\chi^a\Gamma_{ABC}\chi^a\right)^2  
~.~ \,
\end{eqnarray}  
One can also write the transformation of the gravitino fields as
~\footnote{One should notice that
after compactification, the remaining $\psi_A$ where A is the
Calabi-Yau manifold index will be the mixing
of  the $\psi_A$, which are singlets of $SU(3)$ homonomy group. We are
a little sloppy by just writting it as this way}:
\begin{eqnarray}
\delta\psi_A&=&D_A\eta+{\sqrt 2 \over {288}}G_{IJKL}
\left(\Gamma_A^{IJKL}-8\delta_A^I\Gamma^{JKL}\right)\eta
\nonumber\\&&
-{1 \over\displaystyle {1154 \pi^2 \rho_p}} 
{1\over {1-{x\over 3}}}
\left({\kappa \over {4\pi} }\right)^{2/3}
\left(\bar\chi^a\Gamma_{BCD}\chi^a\right)\left(\Gamma_A^{BCD}-
\right.\nonumber\\&&\left.
6\delta_A^B
\Gamma^{CD}\right)\eta+\ldots
~,~ \,
\end{eqnarray}
\begin{eqnarray}
\delta\psi_{11}&=&D_{11}\eta+{\sqrt 2 \over {288}} G_{IJKL}\left(
\Gamma_{11}^{IJKL}-8\delta_{11}^I\Gamma^{JKL}\right)\eta
\nonumber\\&&
+
{1 \over\displaystyle {1154 \pi^2 \rho_p}} 
{1\over {1-{x\over 3}}}
\left({\kappa \over {4\pi}}\right)^{2/3}
\left(\bar\chi^a\Gamma_{ABC}\chi^a\right)
\nonumber\\&&
\Gamma^{ABC}\eta+
\ldots
~,~ \,
\end{eqnarray}
define a modified field 
strength $\tilde G_{IJKL}$ by 
\begin{eqnarray}
\tilde G_{ABC\,11}&=&G_{ABC\,11}-{\sqrt 2 \over {32\pi^2 \rho_p}}
{1\over {1-{x\over 3}}}
\left({\kappa \over {4\pi}}\right)^{2/3}\bar\chi^a\Gamma_{ABC}
\chi^a
~,~ \,
\end{eqnarray}
\begin{eqnarray}
\tilde G_{ABCD}&=&G_{ABCD}
~,~ \,
\end{eqnarray}
one can write the above supersymmetry transformation of the
gravitino field in the hidden sector and in the 4-dimension as:
\begin{eqnarray}
\delta\psi_A&=&D_A\eta+
{\sqrt 2 \over {288}}\tilde G_{IJKL}\left(\Gamma_A{}^{IJKL}-8\delta_A^I
\Gamma^{JKL}\right)\eta+\ldots
~,~ \,
\end{eqnarray}
\begin{eqnarray}
\delta\psi_{11}&=&D_{11}\eta+{\sqrt 2 \over {288}}
\tilde G_{IJKL}\left(\Gamma_{11}{}^{IJKL}-8\delta_{11}^I
\Gamma^{JKL}\right)\eta
\nonumber\\&&
+{1 \over\displaystyle {192\pi^2 \rho_p}}
{1\over {1-{x\over 3}}}
\left({\kappa\over {4\pi}}\right)^{2/3}
\left(\bar\chi^a\Gamma_{ABC}\chi^a\right)\Gamma^{ABC}\eta+\ldots
~.~ \,
\end{eqnarray}
Therefore, as previously, the supersymmetry variations of
$\psi_A$ do not include gaugino condensation term, but, 
the supersymmetry variations of
$\psi_{11}$ do include gaugino condensation term. The 
$\psi_{11}$ will play the role of the goldstino in this scenario.

Now, let us calculate the gravitino mass by gaugino condensation~\cite{NOY}.
 The 
term which is relevant 
in the eleven dimension Lagrangian is:
\begin{eqnarray}
-{\sqrt 2 \over {384 \kappa^2}}\left(\bar\psi_I\Gamma^{IJKLMN}\psi_N
+12\bar\psi^J\Gamma^{KL}\psi^M\right)\left(G_{JKLM}
+\hat G_{JKLM}\right) ~,~\,
\end{eqnarray}
after compactification and considering gaugino condensation, we
obtain:
\begin{eqnarray}
M_{3/2}&=& {1\over\displaystyle {64\pi \alpha_{GUT}}} 
{{1-x}\over\displaystyle {(1+x)(1-{x\over 3})}}
{{\Lambda^3} \over {M_{pl}^2}}
~,~\,
\end{eqnarray}
where $\Lambda$ is defined by following equation:
\begin{eqnarray}
<\bar \chi^a \Gamma_{ijk} \chi^a> &=&
\Lambda^3 \epsilon_{ijk}
 ~.~\,
\end{eqnarray} 

It has been argued that $x$ can not be larger than 0.97 in order to
let the hidden sector physical Calabi-Yau  mass scale ( $M_H$ ) less than
the eleven dimension Planck scale $M_{11}$. In fact, if $M_H > M_{11}$,
one need to worry about the anomaly at the scale between $M_H$ and $M_{11}$.
Here, if we consider the gaugino condesation SUSY breaking,
we can have another reason that we can not let $x =  1$, because we
will have massless gravitino.

Now, let us discuss the gaugino condensation scale. First, taking 
$\alpha_{GUT} =0.04$ and consider the standard embedding $x > 0$.
For $M_{3/2}=100$ GeV, we obtain that:
\begin{eqnarray}
\Lambda = \left({{(1+x)(1-{x \over 3})} \over\displaystyle
{1-x}}\right)^{1/3} \times 1.67 \times 10^{13} GeV
~,~ \,
\end{eqnarray}
for $M_{3/2}=1$ TeV, we obtain:
\begin{eqnarray}
\Lambda = \left({{(1+x)(1-{x \over 3})} \over\displaystyle
{1-x}}\right)^{1/3} \times 3.6 \times 10^{13} GeV
~,~\,
\end{eqnarray}
because x is smaller than 0.97, the factor
\begin{eqnarray}
\left({{(1+x)(1-{x \over 3})} \over\displaystyle
{1-x}}\right)^{1/3} ~,~\,
\end{eqnarray}
will be larger than 1 and smaller than 6.67.
If we consider the case that the gaugino condensation occurs
just at the physical Calabi-Yau compactification scale
in the hidden sector, i. e.,
we consider intermediate unification with gaugino condensation
SUSY breaking, we can obtain that in order to keep gravitino mass
at hundreds GeV range, $\Lambda$  is about   
1.1-2.4$\times 10^{14}$ GeV. 
$M_{11}$ and $M_H$ will be about the same as that scale, and 
$M_{GUT}$ will be about  one half of that
scale. Of course, we can put the realistic GUT scale $M_{GUT}^r$
in the range $10^{12}$ GeV, which is the right unification scale for
$SU(4)\times 
SU(2)_L\times SU(2)_R$ model~\cite{GKLNDT}, 
 and this model has no
problem on proton decay.

Second, taking $\alpha_{GUT} =0.15$ and consider the 
non-standard embedding $x < 0$.
For $M_{3/2}=100$ GeV, we obtain that:
\begin{eqnarray}
\Lambda = \left({{(1+x)(1-{x \over 3})} \over\displaystyle
{1-x}}\right)^{1/3} \times 1.62 \times 10^{13} GeV
~,~\,
\end{eqnarray}
for $M_{3/2}=1$ TeV, we obtain:
\begin{eqnarray}
\Lambda = \left({{(1+x)(1-{x \over 3})} \over\displaystyle
{1-x}}\right)^{1/3} \times 3.5 \times 10^{13} GeV
~,~ \,
\end{eqnarray}
so, the variation is very small. Noticing that the
$\Lambda$ is related to the non-perturbative superpotential as
~\cite{MDRSEW}:
\begin{eqnarray}
\Lambda^3 \sim M_H^3 exp(-{{8 \pi^2} \over {C_2(Q)}} S (1-x)
~,~ \,
\end{eqnarray}
we can not let x close to -1. Therefore, we obtain in this case,
the gaugino condesation scale will occur at $10^{13}$ GeV order.
And if one consider the intermediate unification, 
$M_{GUT}^r$ will be in the range of $10^{11}-10^{13}$ GeV. 

We would like to comment on the superpartner of the goldstino~\cite{NOY}.
Recalling the eleven dimension transformation of the supergravity
multiplet~\cite{HW}:
\begin{eqnarray}
\delta e^m_I= {1\over 2} \bar \eta \Gamma^m \psi_I
 ~,~\,
\end{eqnarray} 
\begin{eqnarray}
\delta C_{IJK} = -{\sqrt 2 \over 8} \bar \eta \Gamma_{\left[IJ
\psi K\right]}
 ~,~\,
\end{eqnarray} 
\begin{eqnarray}
\delta \psi_I = D_I \eta + {\sqrt 2 \over {288}}  
\left(\Gamma_I^{JKL} - 8 \delta_I^J \Gamma^{KLM} \right)
\eta G_{JKLM} + \ldots
 ~,~\,
\end{eqnarray} 
and noticing that in the simplest compactification
and non-deformed Calabi-Yau manifold, the eleven dimension metric can
be written as~\cite{TIAN}:
\begin{eqnarray}
g_{\mu \nu}^{11} = e^{-\gamma} e^{-2\sigma} g_{\mu \nu}^4
 ~,~\,
\end{eqnarray}     
\begin{eqnarray}
g_{i \bar j}^{11} =  e^{\sigma} g_{i \bar j}
~;~ g_{ 11,11}^{11} = e^{2\gamma} e^{-2\sigma} 
 ~.~\,
\end{eqnarray} 
And $S$ and $T$ are defined by:
\begin{eqnarray}
S=e^{3\sigma}+ i 24 \sqrt 2 D
 ~,~\,
\end{eqnarray}     
\begin{eqnarray}
T=e^{\gamma}-i6\sqrt 2 C_5 + |C|^2
 ~,~\,
\end{eqnarray}
where $D$ and $C_5$ are defined by:
\begin{eqnarray}
{1\over {4!}}e^{6\sigma} G_{11 \mu \nu \rho} &=&
\epsilon_{\mu \nu \rho \delta} (\partial^{\delta} D )
 ~,~\,
\end{eqnarray}     
\begin{eqnarray}
C_{5i\bar j} &=& i C_5 \delta_{i \bar j}
 ~.~\,
\end{eqnarray}
Therefore, 
elfbein $e_{11, 11}$ is a function of $S$ and $T$ and
the $C_{IJ 11}$ are the imaginary parts of $S$ and $T$ 
in 4 dimension,
we may conclude that the SUSY might be broken by F-term of $S$ and $T$
~\footnote{One of $F^S$ and $F^T$ might be zero.}.
In addition, if one consider deformed Calabi-Yau manifold, 
$S$ and $T$ are mixed with each other in the two boundaries
~\cite{NOY, LOD}, so, 
SUSY in this case will definitly be broken by the F-term of $S$ and $T$,
although there exist the possibility that one of $F^S$ and $F^T$ might be zero.

\section{Comments on General K\"ahler Function and Superpotential}

When one mentions the difference between the M-theory on $S^1/Z_2$
and the weakly coupled heterotic string, one always points out the
next order correction is large. But, in fact, if the next order correction
is  large, one need to consider the higher order terms ( oder of 
$x^n$ for n $>$ 1).  Therefore, for the simplest compactification, we have
the following general K\"ahler potentil, gauge kinetic
function and non-perturbative superpotential as~\cite{JUNJUN}:
\begin{eqnarray}
K &=& \hat K + \tilde K |C|^2 ~,~ \,
\end{eqnarray}
\begin{eqnarray}
\hat K &=&  -\ln\,[S+\bar S]-3\ln\,[T+\bar T] ~,~\,
\end{eqnarray}
\begin{eqnarray}
\tilde K &=& \left(1+\sum_{i=1}^{\infty} c_i 
\left({{\alpha (T+\bar T)} \over\displaystyle {S+\bar S}} \right)^i
\right)
\nonumber\\&&
({3\over\displaystyle {T+\bar T}})
 |C|^2  ~,~ \,
\end{eqnarray}
\begin{eqnarray}
Ref^O_{\alpha \beta} &=& ReS
\left(1+\sum_{i=1}^{\infty} d_i 
\left({{\alpha T} \over\displaystyle {S}} \right)^i
\right)\, \delta_{\alpha \beta} ~,~\,
\end{eqnarray}
\begin{eqnarray}
Ref^H_{\alpha \beta} &=& ReS
\left(1+\sum_{i=1}^{\infty} d_i 
\left({{-\alpha T} \over\displaystyle {S}} \right)^i
\right)\, \delta_{\alpha \beta} ~,~\,
\end{eqnarray}
\begin{eqnarray}
W_{np} &=& h~ exp(-{{8 \pi^2}\over\displaystyle C_2(Q)}
f^H) ~.~\,
\end{eqnarray} 
So, if x is large, the higher order correction will also be large,
which will change the soft terms, and then change the low energy
phenomenology by RGE running. 

Using standard method~\cite{ABIM, VSKJL},
 one can easily calculate the soft terms. And one need
reconsider the 4-dimension Planck scale's expressions,
and also reconsider multiple moduli case, etc.. The detail of those
discussions, will appear elsewhere~\cite{JUNJUN}.

\section*{Acknowledgments}
We would like to thank T. Han and R. J. Zhang for reading the manuscript
and comments.
This research was supported in part by the U.S.~Department of Energy under
 Grant No.~DE-FG02-95ER40896 and in part by the University of Wisconsin 
 Research Committee with funds granted by the Wisconsin Alumni
  Research Foundation.

\begin{figure}
\centerline{\psfig{file=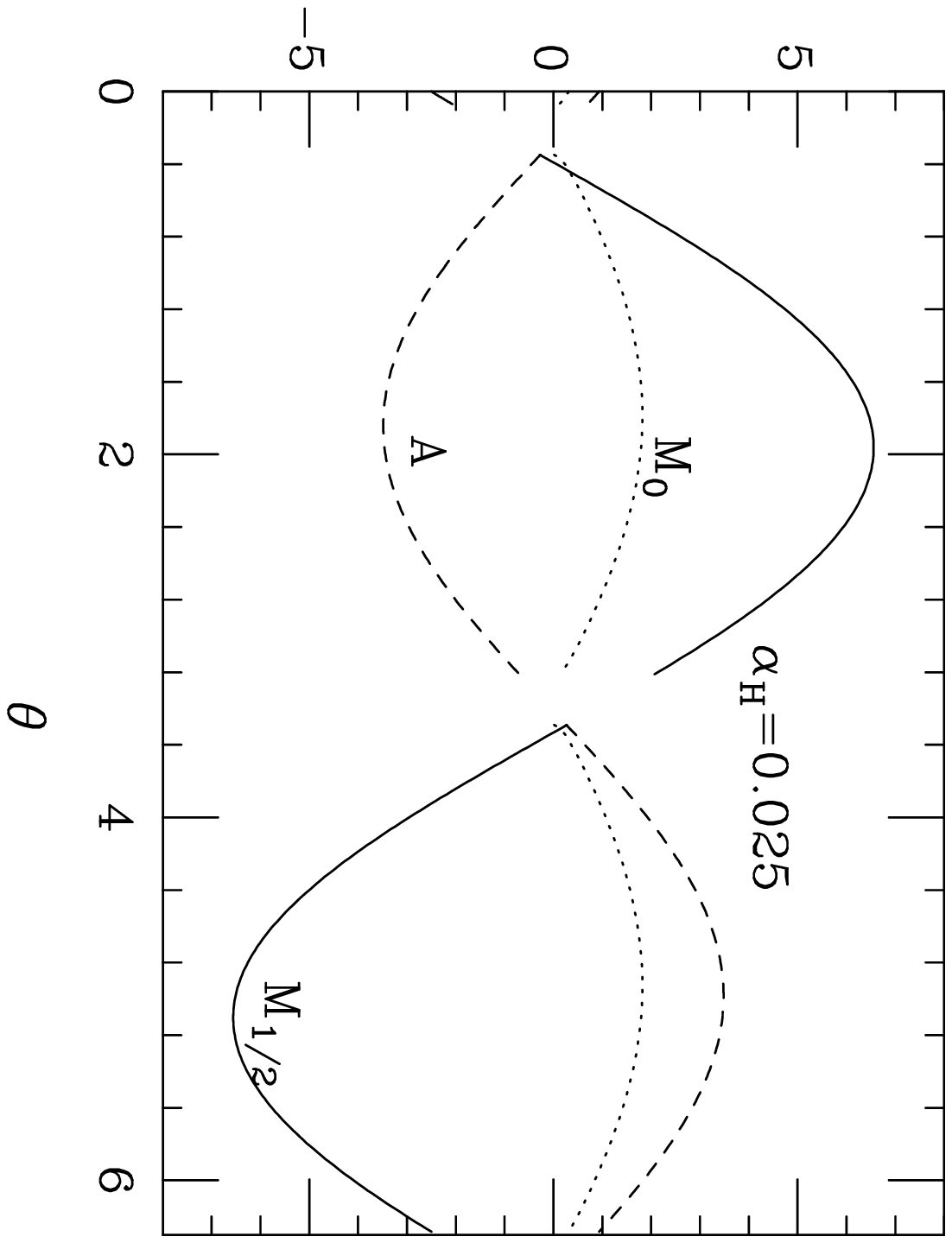,width=15cm}}
\bigskip
\caption[]{Soft terms versus angle $\theta$ with $\alpha_H$=0.025 in
the unit of gravitino mass.}
\label{diagrams}
\end{figure}

\begin{figure}
\centerline{\psfig{file=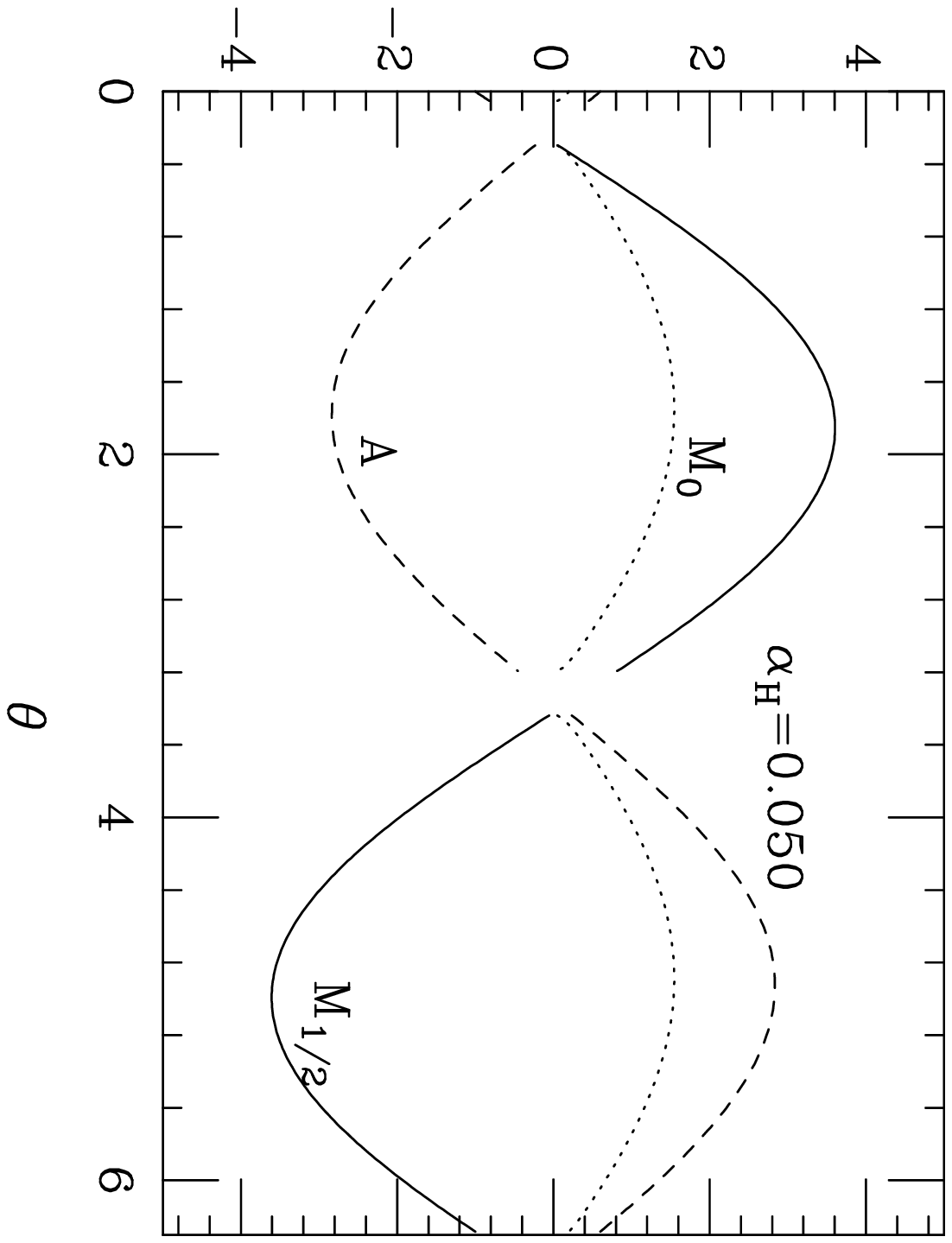,width=15cm}}
\bigskip
\caption[]{Soft terms versus angle $\theta$ with $\alpha_H$=0.05 in
the unit of gravitino mass.}
\label{diagrams}
\end{figure}

\begin{figure}
\centerline{\psfig{file=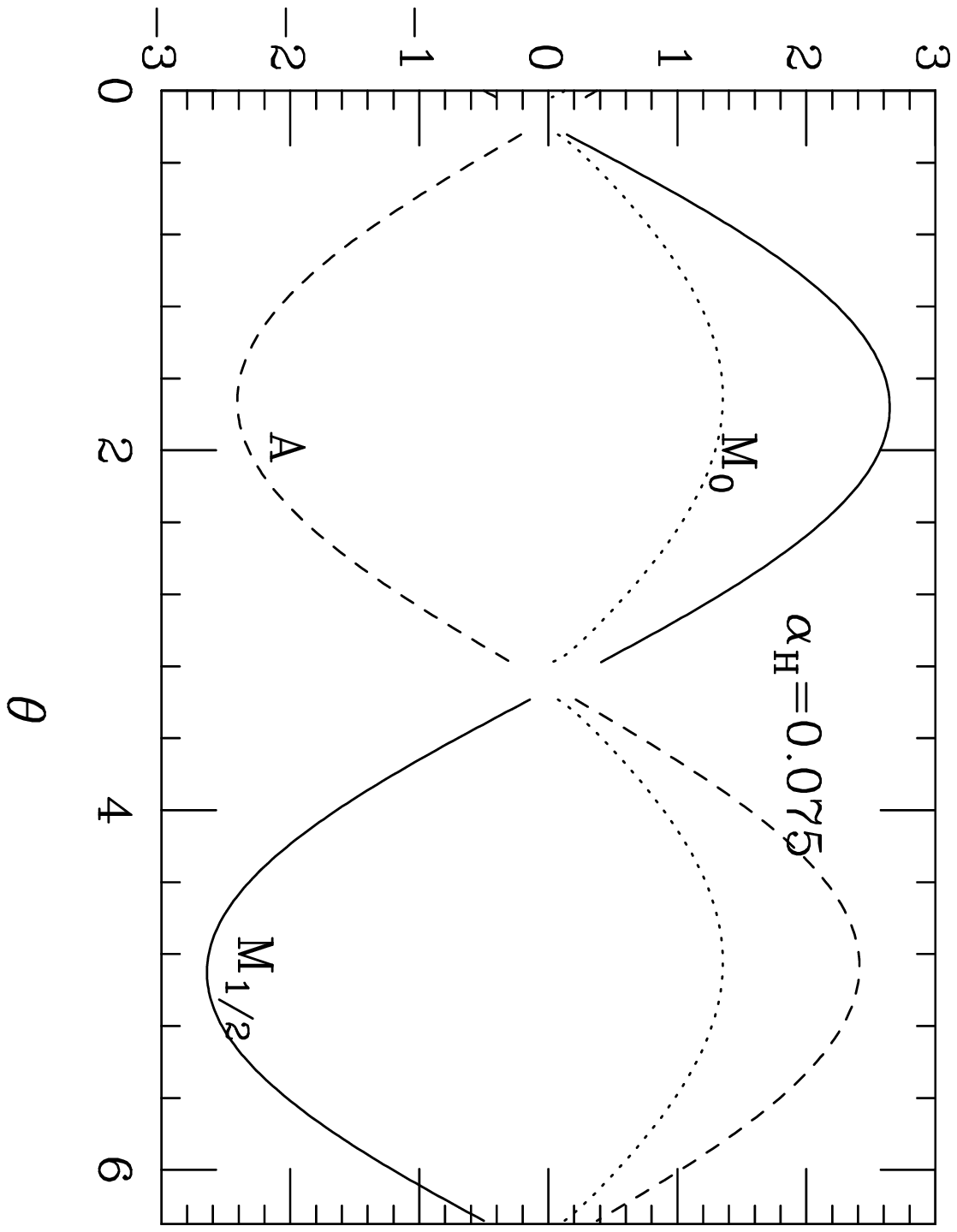,width=15cm}}
\bigskip
\caption[]{Soft terms versus angle $\theta$ with $\alpha_H$=0.075 in
the unit of gravitino mass.}
\label{diagrams}
\end{figure}

\begin{figure}
\centerline{\psfig{file=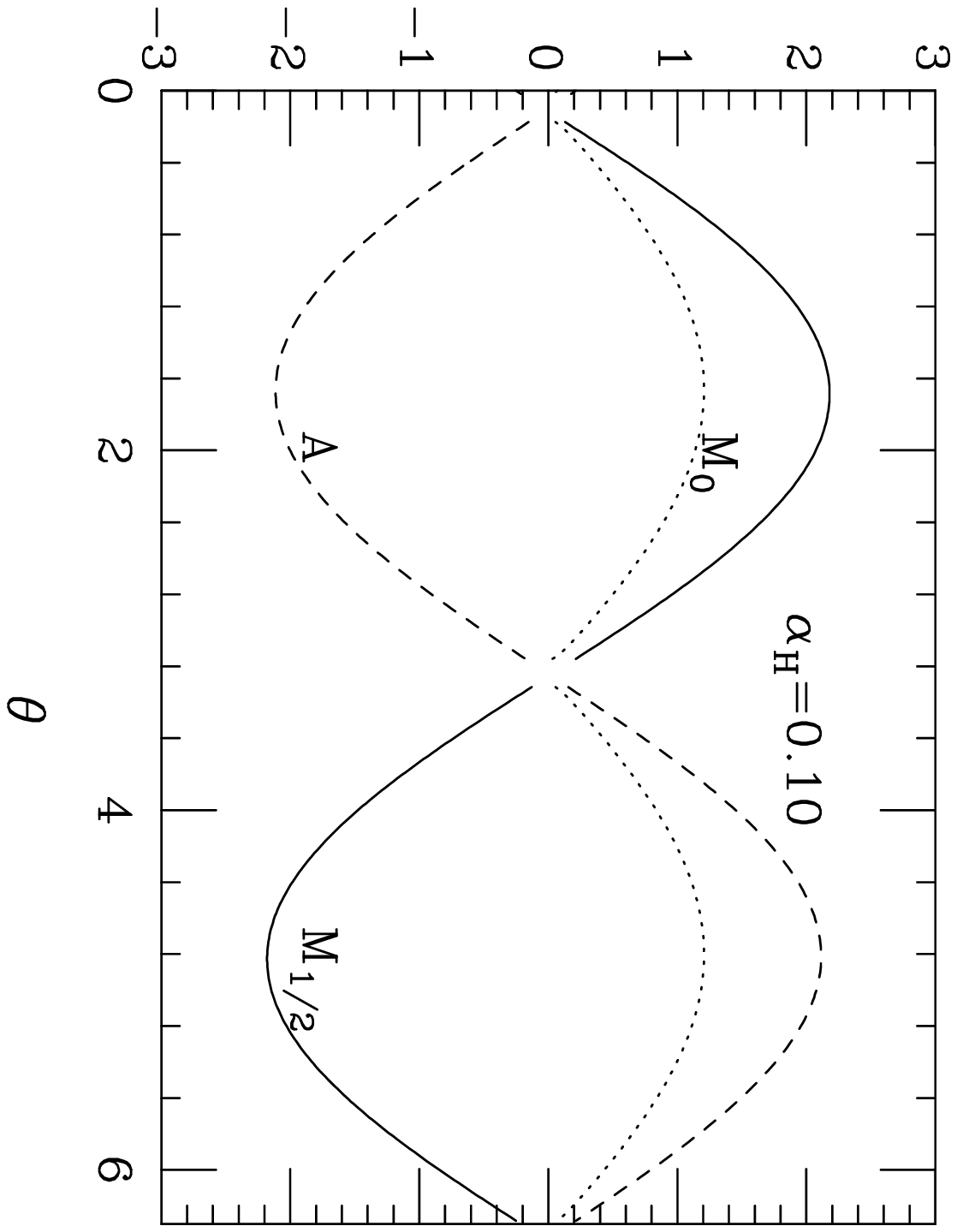,width=15cm}}
\bigskip
\caption[]{Soft terms versus angle $\theta$ with $\alpha_H$=0.1 in
the unit of gravitino mass.}
\label{diagrams}
\end{figure}

\begin{figure}
\centerline{\psfig{file=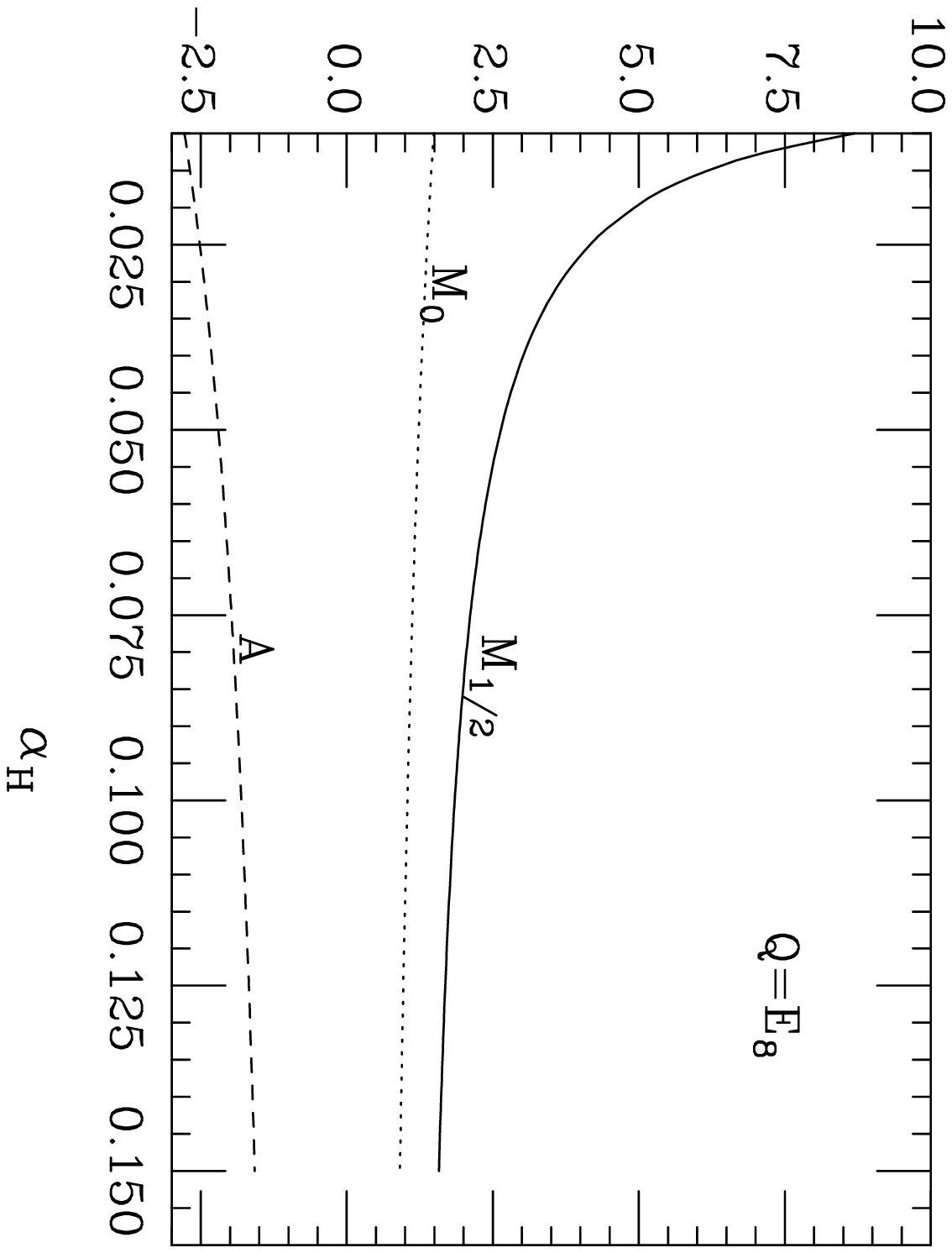,width=15cm}}
\bigskip
\caption[]{Soft terms versus angle $\alpha_H$ with $Q=E_8$ in
the unit of gravitino mass.}
\label{diagrams}
\end{figure}

\begin{figure}
\centerline{\psfig{file=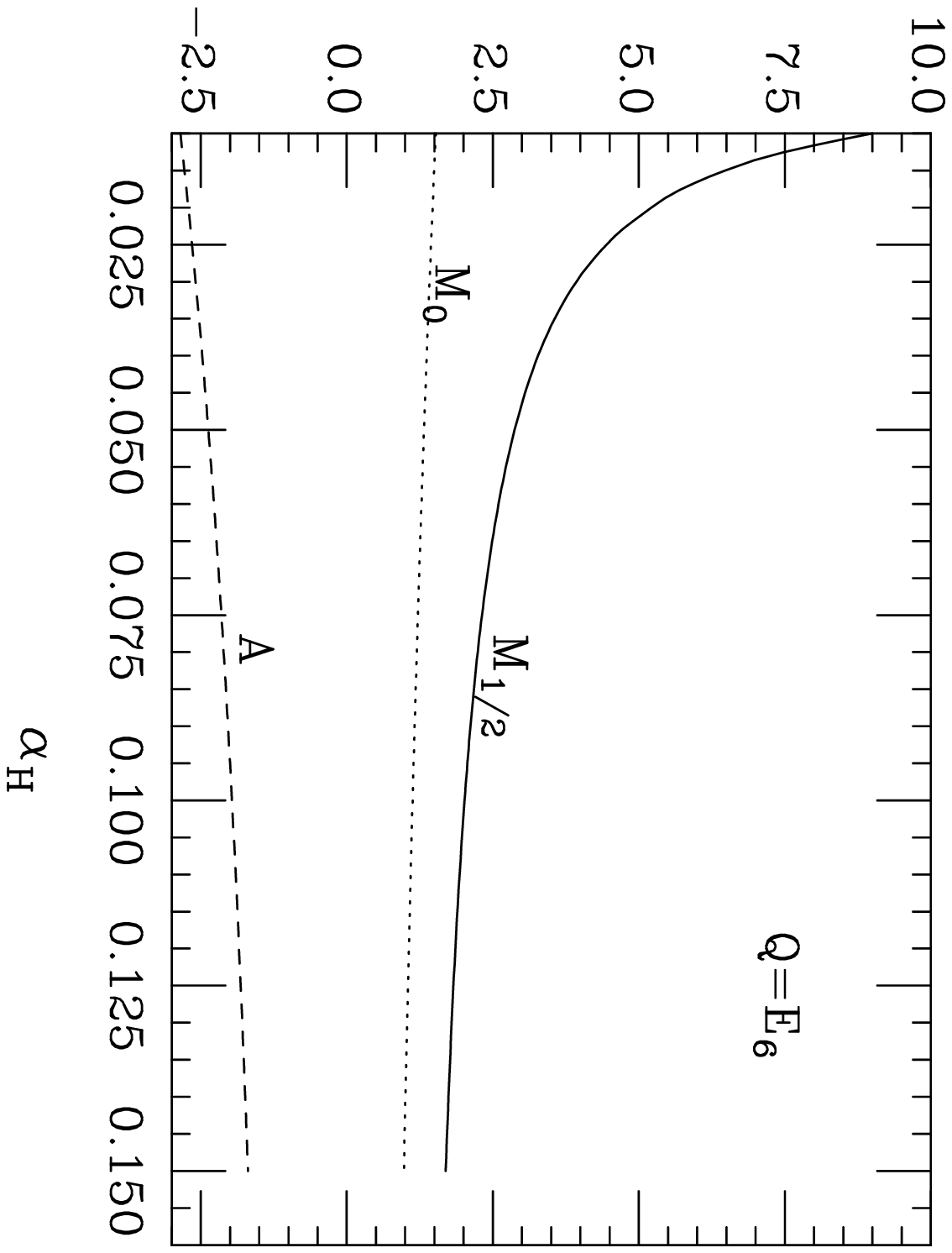,width=15cm}}
\bigskip
\caption[]{Soft terms versus angle $\alpha_H$ with $Q=E_6$ in
the unit of gravitino mass.}
\label{diagrams}
\end{figure}


\newpage

\begin{thebibliography}{99}
\itemsep 0.5mm
\bibitem{HW} P. Horava and E. Witten, Nucl. Phys. B {\bf475} (1996) 94.
\bibitem{Witten} E. Witten, Nucl. Phys. B {\bf471} (1996) 135.
\bibitem{Horava} P. Horava, Phys. Rev. D {\bf54} (1996) 7561.
\bibitem{BD} T. Banks and M. Dine, Nucl. Phys. B {\bf479} (1996) 173 and
hep-th/9609046.
\bibitem{KC} K. Choi, Phys. Rev. D{\bf 56} (1997) 6588;
K. Choi, H. B. Kim, H. Kim, hep-th/9808122.
\bibitem{AQ} I. Antoniadis and M. Quiros, Phys. Lett. B {\bf392} (1997) 61.
\bibitem{AQR} I. Antoniadis and M. Quiros, hep-th/9705037,
hep-th/9707208, hep-th/9709023.
\bibitem{EDCJ} E. Dudas and J. C. Grojean, hep-th/9704177;
 E. Dudas, hep-th/9709043.
\bibitem{VK} E. Caceres, V. S. Kaplunovsky and I. M. Mandelberg,
Nucl. Phys. B{\bf 493} (1997) 73.
\bibitem{LLN} T. Li, J. L. Lopez and D. V. Nanopoulos, hep-ph/9702237,
Mod. Phys. Lett. A. {\bf12} (1997) 2647.
\bibitem{TIAN} T. Li, J. L. Lopez and D. V. Nanopoulos,
Phys. Rev. D{\bf56} (1997) 2602.
\bibitem{BL} T. Li, Phys. Rev. D{\bf 57} (1998) 7539.
\bibitem{JUN} T. Li, hep-th/9804243, Phys. Rev. D to appear.
\bibitem{JUNJUN} T. Li, in preparation.
\bibitem{HLYLZH} C. S. Huang, T. Li, W. Liao, Q. S. Yan and
S. H. Zhou, hep-ph/9810412.
\bibitem{NOY} H. P. Nilles, M. Olechowski and M. Yamaguchi, 
hep-th/9707143, Phys. Lett. B {\bf415} 415 (1997) 24; 
hep-th/9801030.
\bibitem{YNOY} Y. Kawamura, H. P. Nilles, M. Olechowski
and M. Yamaguchi, JHEP 9806:008, 1998, hep-ph/9805397.  
\bibitem{LT} Z. Lalak and S. Thomas, hep-th/9707223.
\bibitem{LOD} A. Lukas, B. A. Ovrut and D. Waldram, hep-th/9710208,
hep-th/9803235.
\bibitem{LOW} A. Lukas, B. A. Ovrut and D. Waldram, hep-th/9711197.
\bibitem{CKM} K. Choi, H. B. Kim and C. Munoz, hep-th/9711158.
\bibitem{BKL} D. Bailin, G. V. Kraniotis and A. Love, 
hep-ph/9803274.
\bibitem{EFN} J. Ellis, A. E. Faraggi and D. V. Nanopolous,
hep-th/9709049.
\bibitem{DM} E. Dudas and J. Mourad, hep-th/9701048.
\bibitem{KBENAKLI} K. Benakli, hep-th/9804096.
\bibitem{KARIM} K. Benakli, hep-th/9805181.
\bibitem{KARIMB} K. Benakli, hep-th/9809582.
\bibitem{SST}S. Stieberger, CERN-TH-98-228 (Jul 1998) 34p.
    [HEP-TH 9807124] to appear in Nucl. Phys. B
\bibitem{NSLPT} Z. Lalak, S. Pokorski, S. Thomas,
hep-ph/9807503.
\bibitem{NSLOW} A. Lukas, B. A. Ovrut, D. Waldram,
hep-th/9808101, hep-th/9901017.
\bibitem{LOSW} A. Lukas, B. A. Ovrut, K. S. Stelle and D. Waldram,
hep-th/98003235, hep-th/9806051.
\bibitem{JELPP} J. Ellis, Z. Lalak, S. Pokoski, W. Pokorski,
hep-ph/9805337, hep-th/9811133.
\bibitem{DVN} D. V. Nanopolous, hep-th/9711080.
\bibitem{CIMUNOZ} J. A. Casas, A. Ibarra and C. Munoz,
 hep-ph/9810266. 
\bibitem{JEDVN} J. Ellis, P. Kanti, N. E. Mavromatos 
and D. V. Nanopolous, hep-th/9711163.
\bibitem{DLOW} R. Donagi, A. Lukas, B. A. Ovrut, D.  Waldram,
 hep-th/9811168, hep-th/9901009.
\bibitem{EXTRAD} For example: 
I. Antoniadis, Phys. Lett. B{\bf 246} (1990) 377.
N. Arkani-Hamed, S. Dimopoulos,
and G. Dvali, Phys. Lett. B{\bf 429} (1998) 263;
N. Arkani-Hamed, S. Dimopoulos, J. March-Russell, hep-th/9809124.
I. Antoniadis, N. Arkani-Hamed, S. Dimopoulos,
and G. Dvali, Phys. Lett. B{\bf 436} (1998) 257.
K. Dienes, E. Dudas, T. Gherghetta, hep-ph/9803466,
hep-ph/9806292, hep-ph/9807552. J. D. Lykken, hep-th/9603133.
T. Han, J. D. Lykken, R. J. Zhang, hep-ph/9811350.
\bibitem{CPBIQ} C. P. Burgess, L. E. Ibanez, F. Quevedo, 
hep-ph/9810533.
\bibitem{GKLNDT} G. K. Leontaris, N. D. Tracas, hep-ph/9902368.
\bibitem{MDRSEW} M. Dine, R. Rohm, N. Seiberg, E. Witten,
Phys. Lett. B{\bf 156} (1985) 55.
\bibitem{ABIM} A. Brignole, I. E. Ibanez and C. Munoz,
hep-ph/9707209; Nucl. Phys. B {\bf 422} (1994) 125.
\bibitem{VSKJL} V. S. Kaplunowsky and J. Louis, Phys. Lett.
B{\bf 306} (1993) 269.
\end{thebibliography}
\end{document}